\documentclass[fleqn,usenatbib]{mnras}
\usepackage{newtxtext,newtxmath}
 \usepackage[T1]{fontenc}
\DeclareRobustCommand{\VAN}[3]{#2}
\let\VANthebibliography\thebibliography
\def\thebibliography{\DeclareRobustCommand{\VAN}[3]{##3}\VANthebibliography}
\def\paren#1{\left( #1 \right)}

\def\angle#1{\left\langle #1 \right\rangle}
\def\ltsima{$\; \buildrel < \over \sim \;$}
\def\lsim{\lower.5ex\hbox{\ltsima}}
\def\gtsima{$\; \buildrel > \over \sim \;$}
\def\gsim{\lower.5ex\hbox{\gtsima}}
\usepackage{graphicx}	
\usepackage{float}
\usepackage{amsmath}	
\usepackage{subfigure}
\usepackage{orcidlink}

\title[Probing GRB jet symmetry via afterglow polarimetry]{Probing the axial symmetry of gamma-ray burst
jets using afterglow polarimetry}

\author[T. Baxter and S. Kobayashi]{
Thomas Baxter$^{\orcidlink{0009-0004-4738-9022},}$$^{1}$ and 
Shiho Kobayashi$^{1}$
\\
$^{1}$Astrophysics Research Institute, Liverpool John Moores University, IC2, Liverpool Science Park, 146 Brownlow Hill, Liverpool L3 5RF, UK\\
}

\date{Accepted XXX. Received YYY; in original form ZZZ}

\pubyear{2024}

\begin{document}
\label{firstpage}
\pagerange{\pageref{firstpage}--\pageref{lastpage}}
\maketitle 

\begin{abstract}
Polarisation measurements of gamma-ray burst afterglows provide a powerful tool for probing the structure of relativistic jets. In this study, we revisit polarisation signals observed in gamma-ray burst afterglows, focusing on the effects of non-axisymmetric jet structures.
To characterize these non-axisymmetric jets, we adopt a simple elliptical jet head model and investigate how deviations from axisymmetry influence the temporal evolution of polarisation properties, particularly around the jet break.
Our results show that the polarisation degree curve typically exhibits two peaks for top-hat jets or a single peak for structured jets, even in the presence of an elliptical jet head. In non-axisymmetric jets, a complete drop in polarisation between peaks is generally absent, and the position angle rotation between the peaks can deviate significantly from 90 degrees. In single-peak cases, the polarisation position angle evolves gradually, contrasting with the constant position angle expected in axisymmetric jets.
We also explore the implications of these findings for recent GRB events, including GRB 121024A, GRB 091018, GRB 020813, and GRB 210610B.
\end{abstract}

\section{Introduction}

The afterglow of a gamma-ray burst (GRB) emerges as the relativistic jet decelerates due to its interaction with the surrounding circumburst material. This interaction generates a forward shock that propagates as a collimated blast wave. Electrons accelerated within these shocks produce the observed non-thermal emission via synchrotron radiation \citep{sari_spectra_1998}.   

The simple top-hat jet model has been remarkably successful in explaining many GRB afterglow observations. This success stems from the fact that, for an on-axis observer, the emission from the jet core dominates, resulting in an afterglow light curve that shows no distinct features indicative of the jet's structure. However, the groundbreaking gravitational-wave-triggered detection of a binary neutron star merger, linked to GRB 170817A \citep{abbott_multimessenger_2017}, demonstrated that the afterglows of off-axis events are highly sensitive to the structure of the jet (e.g., \citealt{lamb_electromagnetic_2017}). Observations have shown that the late-time afterglow light curve of GRB 170817A is consistent with Gaussian jet models (\citealt{lazzati_late_2018}; \citealt{margutti_binary_2018}; \citealt{ghirlanda_compact_2019}; \citealt{troja_year_2019}; \citealt{lamb_optical_2019}). 
Meanwhile, a shallow angular profile in relativistic jet models (e.g., power-law jets) has been proposed for GRB 221009A, the brightest burst ever observed (\citealt{lesage_fermigbm_2023}; \citealt{burns_grb_2023}).
The observed monochromatic steepening of the X-ray and optical light curves strongly suggests a geometrical effect, such as a jet break, although the post-break decay rates are shallower than expected. This discrepancy can be resolved if the afterglow emission originates from a structured jet with a shallow angular energy profile (\citealt{oconnor_structured_2023}; \citealt{birenbaum_afterglow_2024}). 

Numerical simulations suggest such jet structures can arise from the interaction between the jet and the confining medium \citep{gottlieb_structure_2021}. Specifically, this occurs as the jet breaks out through the stellar envelope in the collapsar model (e.g. \citealt{zhang_relativistic_2003}; \citealt{lazzati_universal_2005}; \citealt{gottlieb_structure_2021}); or interacts with merger ejecta (e.g. \citealt{perego_neutrinodriven_2014}; \citealt{nativi_are_2022}).
While most afterglow studies to date have focused on axisymmetric jets (either top-hat or structured), some have investigated azimuthal asymmetries in the energy and velocity distributions. An early example is \citealt{nakar_polarization_2004}, who examined polarisation signals in the patchy-shell model, demonstrating that angular inhomogeneities in an expanding shock can lead to light-curve variability, accompanied by continuous, correlated changes in both the polarisation degree and position angle. 
More recently, \citealt{li_characteristics_2023} and \citealt{li_polarization_2024a}, discuss azimuthal energy and velocity inhomogeneities produced by internal non-uniform magnetic dissipation processes or the precession of the central engine \citep{huang_jet_2019}.
Recent three-dimensional hydrodynamic simulations of jets in the aftermath of neutron star mergers (\citealt{gottlieb_jet_2022}; \citealt{lamb_inhomogeneous_2022}) further highlight these complexities. These studies show that when a jet breaks out of the ejecta, the resulting outflow exhibits a power-law-like polar energy distribution with rotational inhomogeneity. Notably, the jet head is not perfectly circular, as often assumed in conventional theoretical models, instead, it tends to exhibit a more elliptical shape. 

Non-axisymmetric or inhomogeneous structures have been primarily discussed in the context of prompt gamma-ray emission (e.g. \citealt{gill_prompt_2024}; \citealt{lazar_gammaray_2009}). 
This emission originates from the original ejecta from the central engine, which carries large-scale magnetic fields generated at the central source. 
In contrast, the magnetic fields in the afterglow-emitting blast wave are thought to originate from shock-driven instabilities, resulting in a highly tangled structure whose properties are largely governed by the anisotropy of the post-shock field. 
 Detailed afterglow light-curve modelling can break degeneracies between jet structure, viewing geometry, and magnetic field anisotropy \citep{gill_constraining_2020}.

Although the mechanisms behind the acceleration and collimation of GRB jets remain uncertain, understanding their structures can provide valuable insights into these processes. This paper explores how jet asymmetries can be constrained through polarisation observations of afterglows. Our study is partly motivated by recent polarisation measurements  of GRB 210610B \citep{aguifernandez_varying_2024}, which revealed a polarisation position angle (PPA) rotation of $\Delta \phi = 54^{\circ}\pm{9}^{\circ}$ around a possible jet break (see section \ref{subsec:210610B} for details). This rotation deviates significantly from $\Delta \phi = 90^{\circ}$  predicted for axisymmetric jets (\citealt{sari_linear_1999}; \citealt{ghisellini_polarization_1999a}).

We aim to investigate how jet axial asymmetry leaves an imprint on the temporal evolution of afterglow polarisation. To break axisymmetry, we model the global jet structure with an elliptical head. 
Jets emerging from a collapsing star are unlikely to naturally form an elliptical head, as this would require a non-axisymmetric stellar structure; in reality, stars are typically axisymmetric around their rotational axis. In double neutron-star mergers, however, the ejecta along the rotational axis originates from the neutron star collision and may display non-axisymmetric features. Nevertheless, it is worth exploring deviations from axisymmetry, since a perfectly circular jet head is itself an idealised case.

We do not claim that GRB jets generally have elliptical heads. Rather, we adopt this geometry as a toy model to explore departures from the standard top-hat jet. While there is no strong physical motivation for explicitly elliptical heads, the eccentricity provides a simple parameter to control the degree of asymmetry in the global jet structure. This serves as a straightforward starting point for investigating non-axisymmetric jets in this context.

We present our model for evaluating polarisation signals around a jet break in Section \ref{sec:pol_behaviour}, where we illustrate the primary polarisation signature of a non-axisymmetric jet.  We describe the distributions of polarisation signals when the line of sight of an observer intersects the jet head (or shock surface) at random locations in Section \ref{sec:characterisation}. 
In Section \ref{sec:non-axi_structured}, we discuss the structured model in the context of non-axisymmetric jets.
In Section \ref{sec:obs} we present case studies. Finally, in Section \ref{sec:conc}, we give our conclusions and discussion.

\section{Polarisation Behaviour at a Jet Break}\label{sec:pol_behaviour}
\begin{figure} 
    \includegraphics[width=\columnwidth]{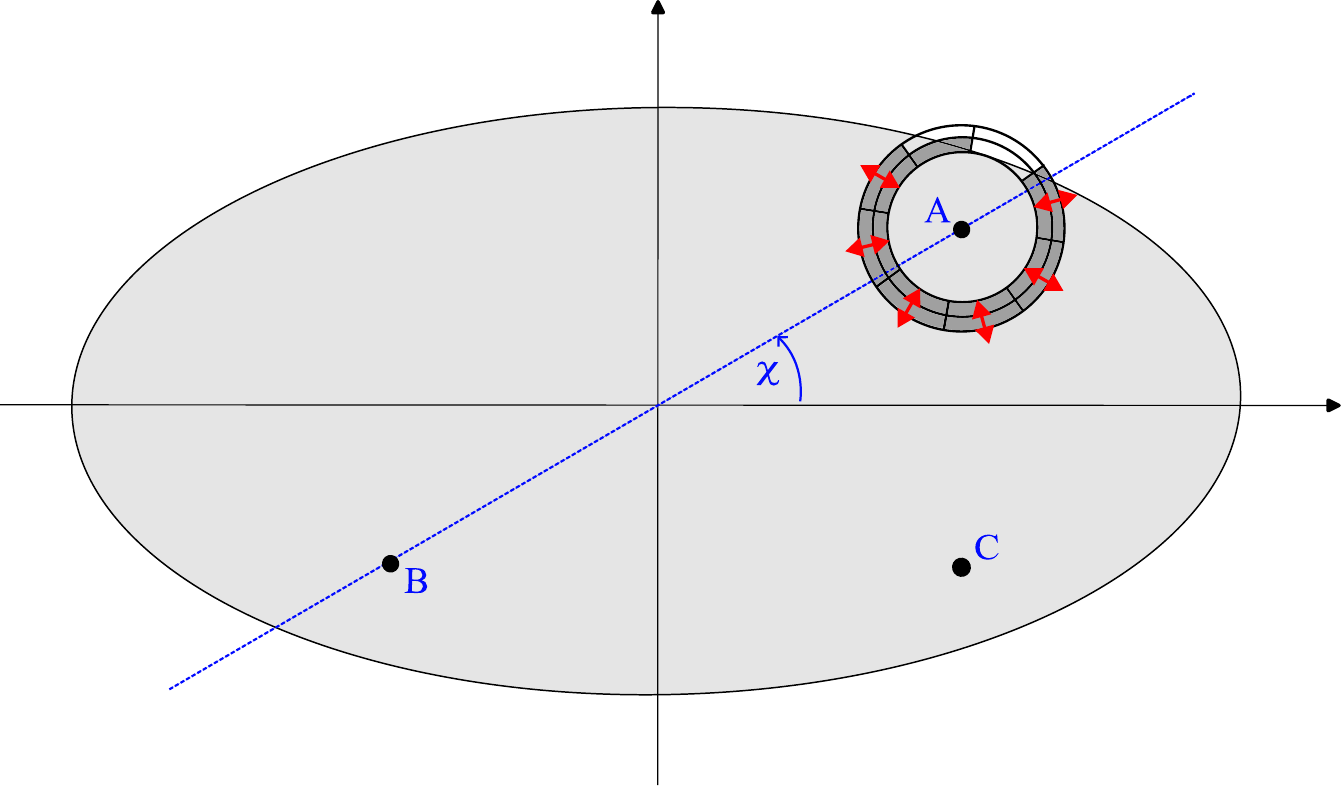}
    \caption{The shaded area represents the physical extent of the jet head, with the line of sight intersecting it at Point A. The location of Point A is specified by two angles: viewing angle $\theta_{view}$, measured from the jet axis (or equivalently expressed as $\xi = (\theta_{view}/\theta_j)$, and the azimuthal angle $\chi$, measured from the semi-major axis of the jet head. The ring surrounding Point A marks the visible region. Red double-headed arrows indicate the polarisation directions of fluid element emission. The point symmetric to Point A with respect to the jet axis (the origin) is Point B, while the point symmetric to Point A with respect to the x-axis is labelled Point C.}\label{fig:geo_model} 
\end{figure}
\begin{figure} 
\includegraphics[width=\columnwidth]{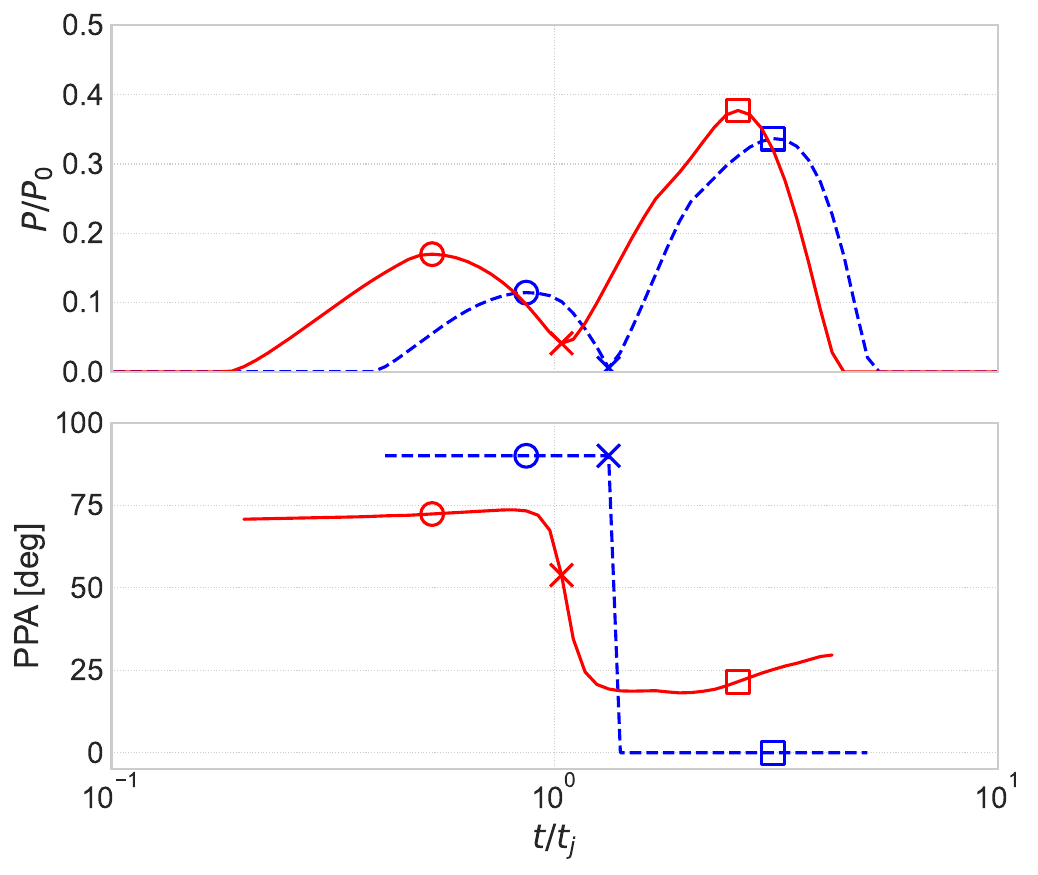}
\caption{Top panel: Polarisation as a function of time,  with both quantities normalised to the the local polarisation degree $P_0$ and the characteristic jet-break time $t_{ j }$}, for $e=0$ (blue dashed) and 0.6 (red solid). 
 The first peaks, the troughs and second peaks in the polarisation light curves are marked by circles, crosses and squares, respectively. The line of sight is characterised by $\chi=1$ and $\xi=0.3$. Bottom panel: Same as the top panel, but showing the the temporal evolution of the polarisation position angle (PPA). 
\label{fig:pcurve}
\end{figure}

In the afterglow phase, magnetic fields in the blast wave are believed to be produced by shock instabilities, though the precise mechanisms responsible for their production in relativistic collisionless shocks remain poorly understood. Since the shock normal (aligned with the radial direction) represents a distinct orientation, the magnetic field components parallel and perpendicular to the shock normal could exhibit significantly different average strengths (\citealt{medvedev_generation_1999}; \citealt{sari_linear_1999}; \citealt{ghisellini_polarization_1999a}).
While many models assume that the post-shock field lies entirely within the shock plane, some observations indicate a more isotropic geometry (\citealt{gill_constraining_2020}; \citealt{stringer_polarization_2020}).

Due to relativistic beaming, an observer at early times can only see a small portion of the jet head’s surface. This visible region is confined to a small area with an angular size of $1/\Gamma$, centred around the point where the line of sight intersects the jet head. Especially, at high frequencies (above the peak synchrotron frequency), most of the emission arises from a ring-like structure, caused by the relativistic limb-brightening effect \citep{granot_images_1999}. Each small segment of the ring might produce highly polarized radiation due to the local anisotropy in the magnetic field, as discussed. However, the overall symmetry of the ring results in a net polarisation of zero.

As noted independently by \citealt{sari_linear_1999} and \citealt{ghisellini_polarization_1999a}, net polarisation signals are expected to emerge around the time of a jet break, when the expanding visible region (i.e., the ring) begins to extend beyond the edge of the jet. We here revisit the polarisation signals for a non-axisymmetric homogeneous (top-hat) jet. 

Following \citealt{sari_linear_1999}, we adopt a toy model to approximate the evolution of the polarisation signal. In this model, (1) the jet head is represented by an ellipse with eccentricity $e$, and the half-opening angle of the jet, particularly along the semi-major axis, is denoted by $\theta_j$. (2) Recent numerical hydrodynamic simulations suggest that lateral jet expansion occurs much more slowly than previously estimated (e.g., \citealt{granot_lateral_2012}). Thus, we assume the lateral expansion is negligible, meaning both the jet opening angle and eccentricity remain constant. The Lorentz factor of the fluid is related to the observer time $t$ by $t/t_j=(\theta_j \Gamma)^{-8/3}$ where $t_j$ is the characteristic time for a jet break.  (3) The (main) visible region is modelled as a thin ring with a radius of $\Gamma^{-1}$, centred on the point where the line of sight intersects the jet head. The ring's width is set to 30\% of its radius. (4) The emission from each fluid element within the ring is polarized in the local radial direction (i.e., perpendicular to the ring) with a polarisation degree of $P_0$.  This polarisation direction assumes that the magnetic field component parallel to the shock normal dominates over the perpendicular component. If the parallel component were weaker, the Stokes parameters $Q$ and $U$ would change sign in the subsequent discussion. However, the discussion is basically identical. (5) The inner region of the ring, corresponding to a circular area with a radius of 0.7 $\Gamma^{-1}$, emits much dimmer radiation—10\% of the ring’s brightness—and is assumed to be unpolarised. (6) Only the portion of the visible region (the ring and the inner circular area) that overlaps with the jet emits photons, uniformly distributed within each area. The region outside the jet does not contribute any emission. Under these assumptions, the evolution of polarisation over time depends on the location of the observer's line of sight within the jet head, characterized by two parameters: the azimuthal angle from the semi-major, $\chi$, and the offset between the observer's line of sight and the centre of the jet, measured in units of the jet opening angle, $\xi=\theta_{view}/\theta_{j}$. 

To estimate the Stokes parameters: $Q$ and $U$, we divide the visible region (the ring and the inner circle) into many segments. Since we consider synchrotron emission from an optically thin blast wave,  the circular polarisation is zero $V=0$. 
Each segment contributes to $Q$ and $U$ as 
\begin{eqnarray} 
dQ &=& P_e dL \cos\paren{2\phi_e} \\
dU &=& P_e dL \sin\paren{2\phi_e} 
\end{eqnarray}
where the local polarisation degree $P_e=P_0$ when the segment lies within the ring, and $P_e=0$ when the segment is inside the inner circle. The local luminosity, $dL$,  is proportional to the area of the segment while also incorporating the constraints imposed by conditions (3), (5), and (6) in the previous paragraph. The angle $\phi_e$ denotes the angle between the positive x-axis (the semi-major axis) and the segment's position vector, measured relative to the line of sight (Point A in Figure \ref{fig:geo_model}). Considering condition (6), $\Sigma dQ$, $\Sigma dU$ and $\Sigma dL$ are evaluated to estimate the net values of the parameters: $q=\Sigma dQ / \Sigma dL$
and $u=\Sigma dU / \Sigma dL$. The net polarisation degree $P$ and position angle $\phi$ are given by 
\begin{eqnarray}\label{eq:PA}
P&=& \sqrt{q^2+u^2}, \\
\phi&=& \frac{1}{2}\arctan\paren{u/q}.
\end{eqnarray}

Figure \ref{fig:pcurve} illustrates the differences in polarisation signals between axisymmetric jets and non-axisymmetric jets (in this paper, non-axisymmetric jets refer to those with elliptical-shaped heads). For axisymmetric jets, the polarisation degree curve (shown as the blue dashed line in the top panel) exhibits two peaks, with the polarisation dropping to zero between the peaks around the jet break time. As seen in the bottom panel, the polarisation position angle (PPA) $\phi$ rotates precisely by $\pi/2$ before and after the point where polarisation vanishes (\citealt{sari_linear_1999}; \citealt{ghisellini_polarization_1999a}). 
In contrast, jets with elliptical heads generally do not exhibit a complete polarisation drop between the two peaks ($P\ne0$ at the trough). The PPA rotates more smoothly, though it still undergoes rapid changes near the trough (approximately at the jet break time). The change in PPA between the two peaks can  significantly deviate from $\pi/2$; in this example, it is approximately $\Delta \phi = 0.28\pi \sim 50.4$ degrees (it is more convenient to express angles in radians for theoretical discussion, while degrees are preferred in the context of observations. Unless otherwise specified, angles are given in radians). 

\section{The Characterisation of polarisation signals}\label{sec:characterisation}
\begin{figure}	
\includegraphics[width=\columnwidth]{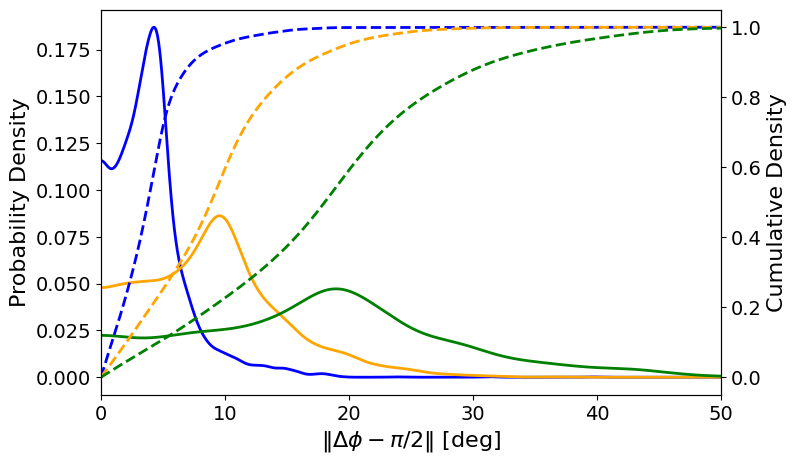}
    \caption{Distribution functions of deviations from $\pi/2$ in the PPA rotation between polarisation peaks for uniformly sampled jets with eccentricities of  $e=0.3$ (blue lines), $e=0.45$ (orange lines), and  $e=0.6$ (green lines). Solid lines represent probability density distributions, while dashed lines show their corresponding cumulative distributions.} 
    \label{fig:CDF_PDF_PArot}
\end{figure}
\begin{figure} 
\centering
\includegraphics[width=\columnwidth]{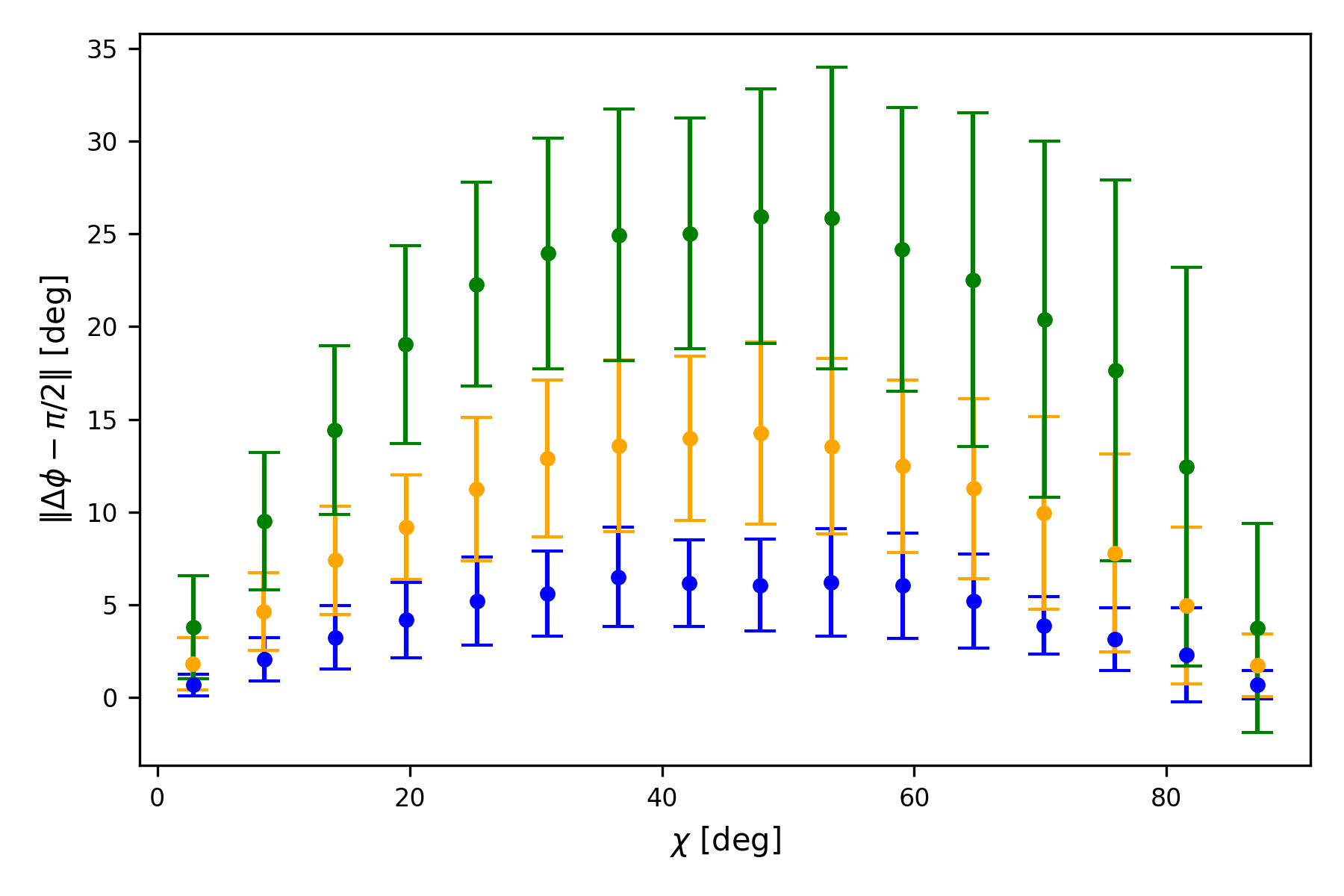}
\caption{The PPA deviations from $\pi/2$ are shown in 16 bins of equal angular width $\chi$ for jets with eccentricities $e=0.3$ (blue), $e=0.45$ (orange), $e=0.6$ (green). The dot represent the averaged values, and the bars indicate the 1 $\sigma$ spreads for each bin. }
\label{fig:dev_dist_chi}
\end{figure}
\begin{figure} 
	\includegraphics[width=\columnwidth]{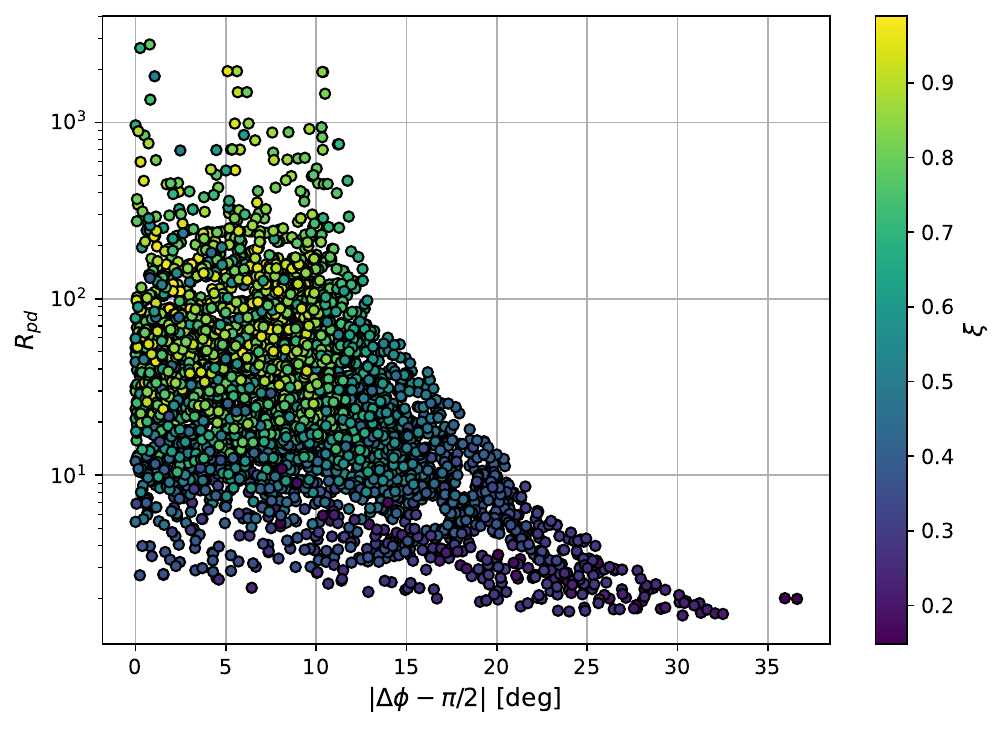}
    \caption{ The deviation of the PPA rotation from $\pi/2$ and 
    the peak-to-trough ratio $R_{pd}$ are shown for the same random sample ($e=0.45$) that generates the orange lines in Figure \ref{fig:CDF_PDF_PArot}. The colour map indicates the offset $\xi$ value of each sample.}
    \label{fig:param_q_ecc_045}
\end{figure}

Consider an observer whose line of sight intersects a jet head, which is characterised by an eccentricity $e$, at the point $(\chi, \xi)$. This intersection point is labelled as Point A in Figure \ref{fig:geo_model}.  
To characterise the time evolution of the PPA, we examine the PPA rotation $\Delta \phi$ between two polarisation peaks and the ratio $R_{pd}=P_{1st}/P_{tr}$, which quantifies the polarisation degrees at the first peak relative to the trough. In Figure \ref{fig:pcurve}, circles and squares denote these polarisation peaks around a jet break. In particular, we analyse $|\Delta \phi - \pi/2|$, which measures deviations from the expected $\pi/2$ rotation in axisymmetric jets. As discussed in section 2, non-axisymmetric jets generally do not exhibit a complete polarisation drop between the two peaks, resulting in lower values of $R_{pd}$.

We investigate the distribution of polarisation signals under the assumption that the intersection points of an observer's line of sight with the jet head are uniformly distributed. Due to the symmetry of the system, it is unnecessary to sample the entire ellipse $(-\pi < \chi < \pi)$; instead, it is sufficient to consider a single quadrant $(0< \chi < \pi/2)$.  

To illustrate this, we first introduce Point B, which is symmetric to Point A with respect to the origin. The distribution of fluid elements around Point B (or the relative position of the jet edge) is identical to that around Point A when rotated by $\pi$ around Point B. Consequently, given the angular dependences in Eqs. (1) and (2), the polarisation signals at any given time are therefore identical for Points A and B. 

Similarly, consider Point C, which is symmetric to Point A with respect to the x-axis. The fluid element distribution around Point C mirrors that of Point A but with a reflection of $\phi \leftrightarrow -\phi$ (flipping vertically with respect to the line passing through Point A and parallel to the x-axis). While the absolute values of the Stokes parameters at Point C match those at Point A,  one of them flips sign, specifically $(q, -u)$. Nevertheless, the position angle rotation $|\Delta \phi - \pi/2|$ and the ratio $R_{pd}$ remain the same as at Point A.  

To investigate intrinsically non-axisymmetric outflows, we model the asymmetric emitting surface with an elliptical cross section, recovering a conical jet when the eccentricity of the jet head is zero, while assuming that the surface dynamics (i.e., the radial motion of each fluid element) still follow the relativistic expansion described by the Blandford-McKee solution. As a result, the emitting surface possesses curvature.
The ellipse represented in Figure \ref{fig:pcurve} represents the projection of the jet head onto the sky. 
The opening angle of GRB jets or their core size is typically only several degrees. For such small angles,  uniform sampling over projected area is approximately equivalent to uniform sampling over the solid angle of the emitting surface. 

Figure \ref{fig:CDF_PDF_PArot} shows the distributions of $|\Delta \phi - \pi/2|$, derived from 5000 uniform sampled line of sight realisations across a quadrant of the elliptic jet head for eccentricity $e=0.3$, 0.45, and 0.6.
For axisymmetric jets ($e=0$), the probability distribution is given by a delta function at 0, meaning the PPA rotation $\Delta \phi$ is always $\pi/2$.  As eccentricity increases, the probability distribution (solid lines) peaks at larger deviation angles, with the distribution's spread also increasing. The probability density distributions peak at $|\Delta \phi - \pi/2|$ = 4.3, 9.5 and 19 degrees, while their cumulative distributions reach 50\% at 3.8, 8.9, 18 degrees, and 90\% at 7.3, 17, 31 degrees for $e=0.3, 0.45$ and 0.6, respectively.  Since the semi-minor axis is smaller than the semi-major axis by a factor of $\sqrt{1-e^2}$, it is reduced by approximately 5\%--20\% (or equivalently by a factor of $\sim 0.8 -0.95$) across this range of eccentricities. Our results show that even a mild deformation of the jet head leads to PPA rotations that deviate significantly from $\pi/2$. 

When the line of sight intersects the jet head near its semi-major or semi-minor axis, the geometry exhibits greater symmetry, leading to a position angle rotation close to $\pi/2$. As expected, Figure \ref{fig:dev_dist_chi} shows that the deviation from $\pi/2$ increases with higher eccentricity, particularly for intermediate angles ($\chi \sim \pi/4$). The spread in the probability distributions shown in Figure \ref{fig:CDF_PDF_PArot} can be partly attributed to variations in PPA rotation between the more symmetric configurations ($\chi \sim 0$ or $\pi/2$) and the asymmetric configuration ($\chi \sim \pi/4$).  In Figure \ref{fig:dev_dist_chi}, 
 for $e=0.6$, the azimuthal bin nearest the semi-minor axis ($\chi\sim 90 \text{ deg}$ ) results in  negative values for $1\sigma$ error. This is due to the significant positive skewness of the distribution. 

Additionally, the offset $\xi=\theta_{view}/\theta_j$ also influences the rotation of the position angle. Figure 
\ref{fig:param_q_ecc_045} shows $|\Delta \phi - \pi/2|$ and $R_{pd}$ for the same random uniform samples with $e=0.45$, where the colour map represents the offset $\xi$ for each sample.  For a smaller offset (and intermediate $\chi$), the deviation $|\Delta \phi - \pi/2|$ becomes larger (if the line of sight passes regions near the semi-major or semi-minor axis, the rotation is close to $\pi/2$ regardless of the offset values). The high-deviation tails in the probability density distributions (the solid lines), seen above their peaks in Figure \ref{fig:CDF_PDF_PArot}, correspond to low-$\xi$ cases ($\xi \lsim 0.4$) where the line of sight is relatively close to the central jet axis. At small offsets $\xi$, the first peak of the polarisation curve becomes less pronounced, leading to a lower $R_{pd}$. For eccentricity $e=0.3$, 0.45 and 0.6, the top 10\% of cases with the largest $\Delta \phi$ deviation from $\pi/2$ have $R_{pd}$ values below approximately 86, 38, and 22. Although these extreme cases show lower $R_{pd}$ for higher eccentricities,  a jet head with a larger  eccentricity $e$ has a higher probability of producing a significant $\Delta \phi$ deviation from $\pi/2$. As discussed in section \ref{subsec:210610B}, observational measurements of  $R_{pd}$ and $\Delta \phi$ can provide valuable constraints on the eccentricity $e$.

For even smaller offsets, one of the peaks, typically the first, becomes less pronounced.
To determine whether a polarisation curve exhibits two peaks and to locate them, we have implemented a peak-finding algorithm similar to the one discussed in \citealt{li_lognormal_1996} (see Appendix \ref{app:pfa} for more details). 
For approximately, 4\%, 7\% and 11\% of uniform random realisations for the $e=0.3$, 0.45 and 0.6, respectively, the polarisation curves exhibit only a single peak. 
These single-peak cases have been excluded from the statistical analysis of PPA rotation presented in Figs \ref{fig:CDF_PDF_PArot}, \ref{fig:dev_dist_chi} and \ref{fig:param_q_ecc_045}.

Single-peak behaviour can be understood from the simple configuration where the line of sight passes through the jet centre (i.e., the origin; see Figure \ref{fig:geo_model} for the following discussion). The visible region forms a ring centred at this point, with each fluid element emitting polarised light in the local radial direction. For a circular jet head ($e = 0$), the system’s symmetry leads to complete cancellation of the polarisation signal. In contrast, for an eccentric jet head ($e \ne 0$), the initial net polarisation is zero, but as the jet decelerates, the top and bottom parts of the ring, emitting vertically polarised light, extend beyond the jet edge, breaking the symmetry. The net polarisation then becomes dominated by the left and right parts of the ring, whose emission is polarised in the horizontal direction. As deceleration proceeds, these segments continue to dominate, resulting in a constant PPA and a single peak in the polarisation curve.
Although the probability of the line of sight passing exactly through the jet centre is zero, we find that the polarisation curve still exhibits a single peak, accompanied by a rotation of the PPA, when the line of sight lies close to the centre. As the viewing angle increases, a minor bump typically emerges during the rising phase and gradually evolves into a two-peak structure.

\section{Non-axisymmetric structured jet}\label{sec:non-axi_structured}
\begin{figure} 
    \includegraphics[width=\columnwidth]{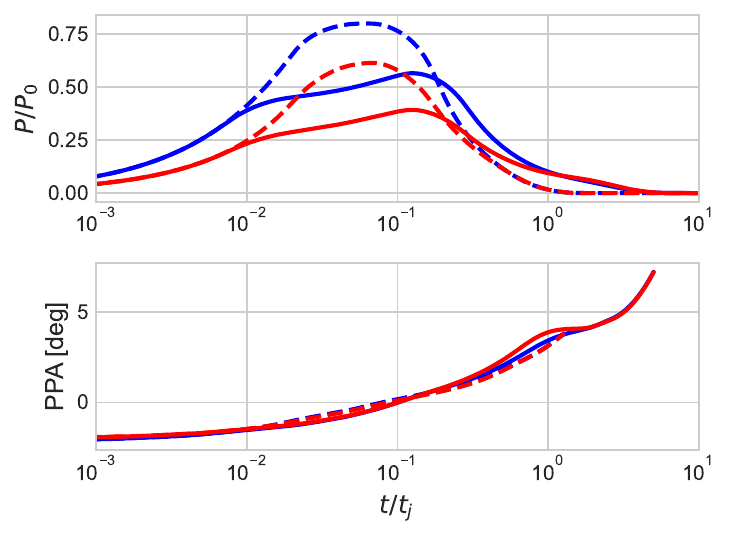}
    \caption{
    Polarisation signals for a line of sight intersecting the outer region, where the surface luminosity density declines following a power-law with an index of $\alpha=2$ (red) or 3 (blue). Two offset angles are considered: $\xi=2\theta_{c}/\theta_{j}$ (solid) and $4\theta_{c}/\theta_{j}$ (dashed).
    Top panel: temporal evolution of polarisation degree.
    Bottom panel: temporal evolution of polarisation position angle}. The model assumes $e=0.3$, $\chi = \pi/4$, and $\theta_{c}/\theta_{j}=0.13$.    
\label{fig:structured_outside_core}
\end{figure}
\begin{figure} 
    \includegraphics[width=\columnwidth]{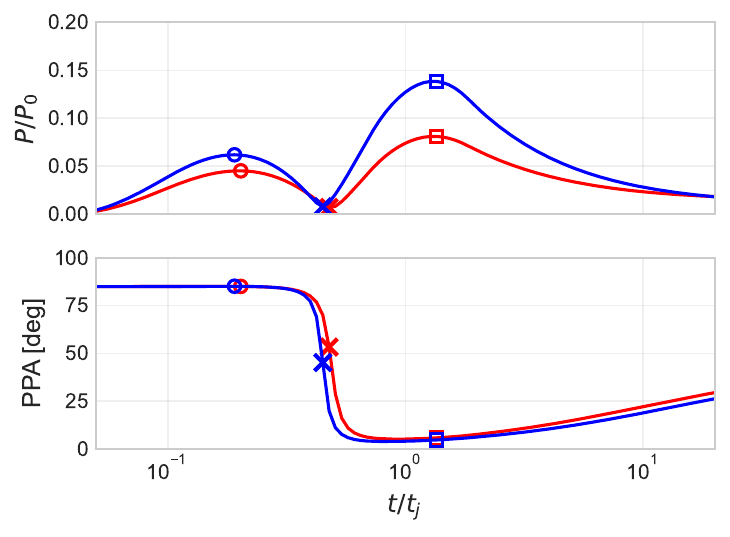}
    \caption{Polarisation signals for a line of sight intersecting the jet core at $\xi=0.5 \theta_{c}/\theta_{j}$. All jet properties are identical to those in Figure \ref{fig:structured_outside_core}, except where specified. Top panel: polarisation degree curves. Bottom panel: Evolution of the polarisation position angle}. Curves are shown for
   $\alpha=2$ (red) and $\alpha=3 $ (blue). The first peaks, the troughs and second peaks in the polarisation light curves are marked by circles, crosses and squares, respectively. 
\label{fig:structured_within_core}
\end{figure}

Observations of GRB 170817A demonstrate that some GRB jets deviate from a simple top-hat structure, instead exhibiting a structured profile. To capture this complexity, we extend our model to incorporate a structured luminosity distribution. In this framework, a uniform central core is encircled by a fainter outer region, where the surface luminosity density gradually declines following a power-law profile, extending out to the jet's maximum opening angle. 

We assume that the core, the jet edge and brightness contours all have a common elliptical shape with a common eccentricity $e$, where the jet core $\theta_c$ and the jet outer edge angles $\theta_j$ are defined along the semi-major axis. Considering the relativistic limb-brightening effect, we adopt the same ring-shaped visible region as in the top-hat jet case. The procedure for estimating polarisation signals remains unchanged, but the local luminosity $dL$ of a segment is now determined by a position-dependent surface luminosity density, which is highest in the core and decreases as a power law with index $\alpha$ in the outer region. 

The luminosity reduction factor for a segment in the outer region, located at $(x,y)$ on the jet head is given by 
\begin{eqnarray}
&\paren{\frac{x^2}{a_c^2}+\frac{y^2}{b_c^2}}^{-\alpha/2}&,
\end{eqnarray}
where $a_c=\theta_c$ and $b_c=\theta_c\sqrt{1-e^2}$ are the semi-major and semi-minor axes of the elliptical jet core.  

In a structured jet, the jet break time for an off-axis observer is determined by the viewing angle $\theta_{view}$, rather than by the jet opening angle as in the top-hat jet model (\citealt{zhang_gammaray_2002}; \citealt{rossi_afterglow_2002}). We estimate the characteristic jet-break time $t_j$ using a simple relation $\Gamma(t_j) \sim 1/\theta_{eff}$, where $\theta_{eff}= \theta_{view}$ if the line of sight is in the outer region, and $\theta_{eff}= \theta_{c}$ if it is within the core.

Figure \ref{fig:structured_outside_core} shows the polarisation signals 
when the line of sight falls in the outer region. As demonstrated by \citealt{rossi_polarization_2004} and \citealt{birenbaum_afterglow_2024},  the polarisation curves differ significantly from those of a top-hat jet, regardless of axial symmetry. 
This discrepancy arises from the non-uniform luminosity distribution in the visible region surrounding the line of sight (i.e., the ring), with the intense emission concentrated toward the jet axis. Consequently, the observed polarisation signal is primarily influenced by these bright segments, resulting in a single peak in the polarisation curve. 

 The position angle evolution also differs from those of a top-hat jet. It consistently aligns with the direction toward the central jet axis (see assumption (4) in Section 2). When the line of sight falls on the jet head near its semi-major or semi-minor axis ($\chi \sim 0$ or $\pi/2$), the geometry becomes more symmetric, the position angle remains constant over time, consistent with the results for axisymmetric structured jets  \citep{rossi_polarization_2004}. However, for a general azimuthal angle $\chi$, the eccentricity of the brightness contour lines leads to a slight evolution of the position angle around the peak of the polarisation degree. This effect is illustrated for $\chi=\pi/4$ in the bottom panel of Figure \ref{fig:structured_outside_core}.  

The PPA rotation over the full width at half maximum of the polarisation peak is evaluated for $\chi=\pi/4$ (the rotation is maximised for $\chi=\pi/4$). 
We obtain rotation values of $\Delta\phi$ = 3.5, 2.9, 2.6 degrees for $e=0.3$, $\Delta\phi$ = 16, 13, 12 degrees for $e=0.6$, corresponding to $\xi/\angle{\theta_{c}/\theta_{j}}=2,3, 4$, respectively. While these values are evaluated for $\alpha=3$, the rotation remains insensitive to $\alpha$. 

As discussed in Section 3, the polarisation degree curve can exhibit a single peak even for top-hat, non-axisymmetric jets when the line of sight is close to the jet axis. In such cases, the PPA typically rotates between a minor bump and the main peak, reaching a local maximum or minimum at the peak of the polarisation curve. The rotation rate of the PPA often differs markedly between the rising and decaying phases. In contrast, for structured jets, the PPA evolves more smoothly and continuously across the peak.

Figure \ref{fig:structured_within_core} illustrates polarisation signals when the line of sight passes through the jet core, corresponding to $\xi=0.5 (\theta_c/\theta_j)$ and $\chi=\pi/4$. The polarisation curves resemble those of top-hat, non-axisymmetric jets (represented by the red solid line in the top panel of Figure \ref{fig:pcurve}), exhibiting two peaks without a complete polarisation dip in between. The temporal evolution of the PPA (shown in the bottom panel) is largely insensitive to the value of $\alpha$. When $\alpha=2$ and 3, we obtain the PPA rotation between the two peaks is $\Delta \phi \sim 79$ degrees and 80 degrees, respectively. In the limit $\alpha \gg 1$ (i.e., the top-hat model), the rotation angle increases slightly to $\Delta \phi \sim 83$ degrees. Overall, the top-hat model serve as a reasonable approximation of the polarisation signals when the line of sight intersects the jet core.

\section{Case Studies}\label{sec:obs}
Optical linear polarisation measurements have been conducted for many GRB afterglows, typically several hours to a few days after the prompt gamma-ray emission (\citealt{covino_polarization_2016}, and references therein). This time frame corresponds to when a jet break is expected to occur.  The observed linear polarisation degree or its upper limits are generally low ($\lsim $ several percent). Such low polarisation levels suggest that 
in shock-generated random magnetic fields, neither the component parallel nor perpendicular to the shock normal dominates entirely (e.g., \citealt{gill_constraining_2020}), though the fields are not fully isotropic. In our model, the local polarisation degree $P_{0}$ is treated as a free parameter that sets the overall normalisation of the polarisation curve but it does not affect its shape or the absolute values and evolution of PPA. Adopting a low value, $P_0 \ll 0.7$ (much smaller than the maximum value for ordered fields), makes the model well-suited for interpreting afterglow polarisation measurements.

\subsection{GRB 020813}\label{subsec:020813}
\begin{figure}
    \centering
    \includegraphics[width=\columnwidth]{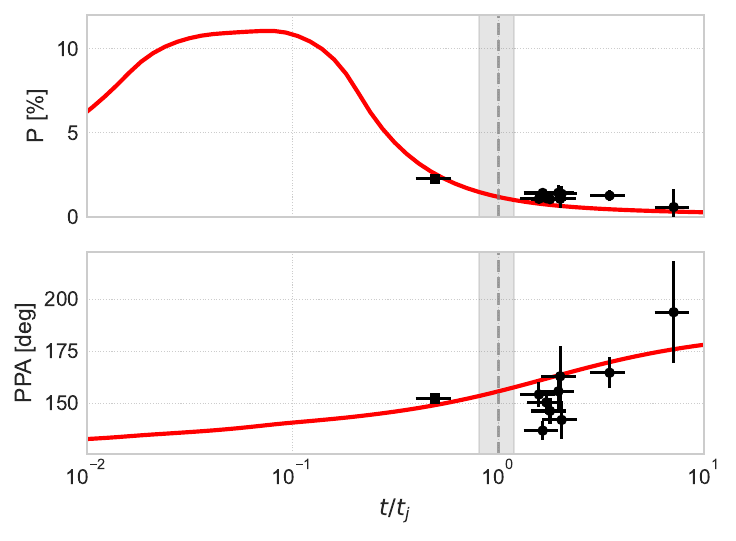}
    \caption{Sample fit for GRB020813 (red solid line). Structured jet model parameters: $e=0.6$, $\theta_{c}/\theta_{j}=0.13$, $\xi=3(\theta_{c}/\theta_{j})$, $\chi=0.8$, 
    and $P_{ 0 }=0.15$.
    Optical polarimetry data from \citep{gorosabel_grb_2004} are shown, where squares (pre-break) and circles represent Keck and VLT measurements, respectively. The time axis is 
  normalised to the break $t_j=0.56 \pm 0.21$ days detected in the V-band light curve, with the break-time uncertainty indicated by the shaded region.}
    \label{fig:GRB020813}
\end{figure}

A bright long-duration burst ($> 125$\,s) detected by HETE-2, located at a redshift of $z=1.255$ \citep{barth_optical_2003}, was studied. 
Polarimetry data obtained with Keck and VLT and \citep{gorosabel_grb_2004} was collected around a jet break, occurring at $\sim 13$ hours post-burst (\citealt{covino_optical_2003}). As shown in Figure \ref{fig:GRB020813}, the data is most consistently described by an almost constant degree of linear polarisation at the $\sim 1\%$ level, along with a stable position angle. While a slow evolution of the position angle $\Delta \phi \sim $ a few tens of degrees cannot be ruled out, the observations do not support a sudden 90 degree rotation of the position angle at the jet break, which is a signature of top-hat axisymmetric jets. Given the sparsity of polarisation data points, a single-peaked polarisation curve remains consistent with the observations. The evolution of both the polarisation degree and position angle can be explained by structured jets, whether axisymmetric or not. If a gradual rotation of the position angle were confirmed, it would indicate a structured, non-axisymmetric jet.

\subsection{GRB 091018}\label{subsec:091018}
\begin{figure}
    \centering
    \includegraphics[width=\columnwidth]{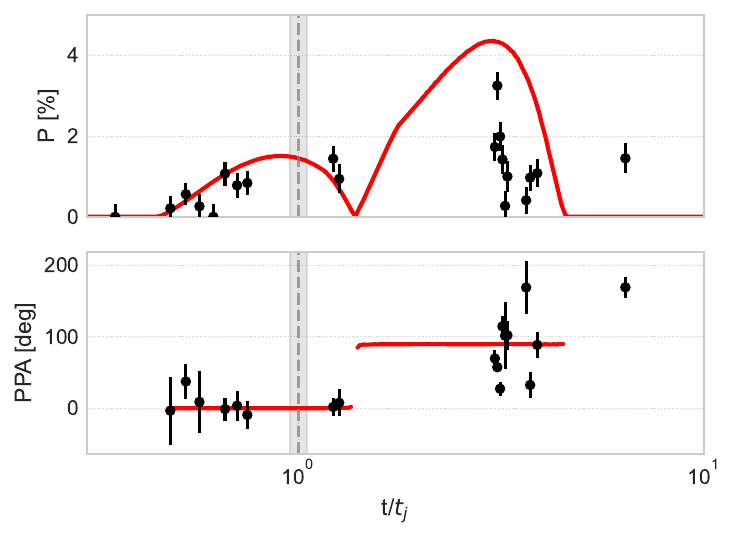}
    \caption{Sample fit for GRB091018 (red solid line). 
    Top-hat jet model parameters: 
    $e=0.15$, $\xi=0.25$, $\chi=-1.48$ and $P_0=0.15$. Optical polarimetry data from \citep{wiersema_detailed_2012}.}
    \label{fig:enter-label}
\end{figure}

This long burst ($T_{90}=4.4$ sec and redshift $z=0.971$) was detected by Swift
\citep{wiersema_detailed_2012}. Its optical light curve is well described by a broken power-law, with a break at $\sim 9$ hours, coinciding with a steepening in the X-ray light curve.  While the post-break decay indices (1.54 in the X-ray and 1.33 in the optical) are shallower than predicted from the standard fireball model, similar trends have been observed in other Swift XRT afterglows, where  potential jet breaks often exhibit relatively shallow decay indices \citep{racusin_jet_2009}. 

Linear polarisation monitoring with VLT \citep{wiersema_detailed_2012} revealed a polarisation curve inconsistent with a single-peaked profile, and instead favouring a two-peak structure. In top-hat jet models, both axisymmetric and non-axisymmetric,  the second peak tends to have a higher amplitude than the first. Although some data points support this trend, significant scatter suggests the presence of a deep, short dip within the second peak. The PPA remains stable and nearly constant during the first peak, but shifts to a higher overall value in the second peak, with significant scatter around the 90-degree rotation.  These variations in both the polarisation degree and position angle during the second peak are not fully explained even by non-axisymmetric jets (i.e., elliptical jet heads). As noted by \citealt{wiersema_detailed_2012}, if the scatter in the second peak is caused by an additional component
such as a bright patch within the jet or microlensing effects that partially cancel the polarisation signals, the presence of a two-peak profile, combined with a nearly constant PPA in the first peak, suggests either a top-hat jet or a view of the core of a structured jet. In this scenario, the jet head (or core) must have a small eccentricity or be observed 
from a line of sight located near its semi-major or semi-minor axis. Additionally, the shallow post-break decay indices may imply that the jet core is encased by an outer region with a shallow angular profile. 

\subsection{GRB 121024A}\label{subsec:121024A}
\begin{figure}
    \centering
    \includegraphics[width=\columnwidth]{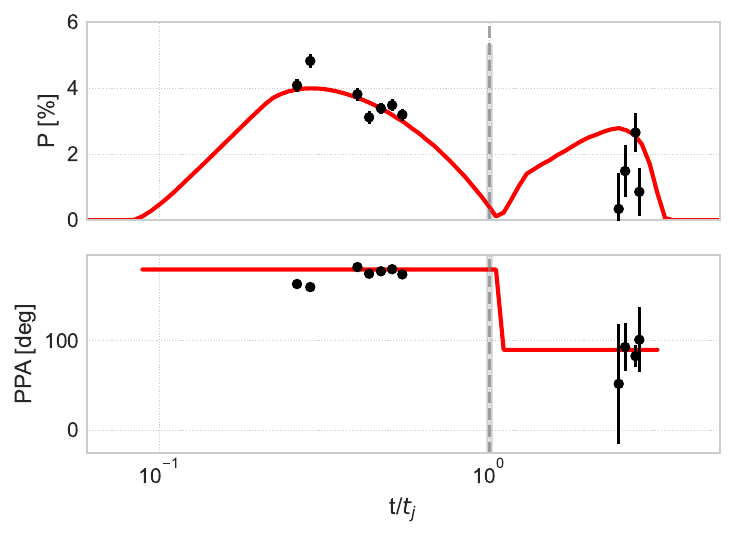}
    \caption{Sample fit of GRB121024A (red solid line). Top-hat jet model parameters: 
    $e=0.65$, $\xi=0.35$, $\chi=\pi/2$ and $P_{ 0 }=0.22$. Optical polarimetry data from \citep{wiersema_circular_2014}.}
    \label{fig:GRB121024A}
\end{figure}

A long burst with $T_{90}=69$ sec was detected by Swift, with a redshift of $z=2.298$ determined shortly afterwards. The X-ray light curve follows a three-segment power-law evolution, with the final break at $t\sim 10$ hours coinciding with a break in the optical light curve \citep{wiersema_circular_2014}.  The monochromatic nature of this break suggests a jet break origin. However, the post-break decay 
indices remain relatively shallow, at 1.67 in the X-rays and 1.25 in the optical. 
VLT polarimetry measurements taken 3-6 hours after the burst show linear polarisation degrees ranging from 3-5\%, with a stable position angle (see Figure \ref{fig:GRB121024A}). However, observations from the following night (after the jet break) reveal a significant shift in the position angle,  consistent with a 90 degree rotation \citep{wiersema_circular_2014}. These polarisation properties also support the similar jet structure interpretations as 
for GRB 091018, although the polarisation degree shows significant scatter (with larger uncertainties) after the break (the top panel), and our model curve does not reproduce these data well in this case either.

\subsection{GRB 210610B}\label{subsec:210610B}
\begin{figure}
    \centering
    \includegraphics[width=\columnwidth]{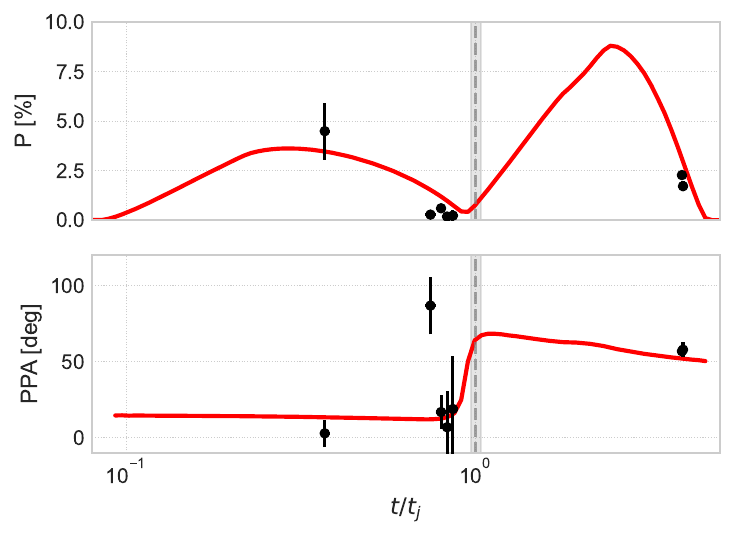}
    \caption{Sample fit of GRB210610B (red solid line). Top-hat jet model parameters: 
    $e=0.7$, $\xi=0.35$, $\chi=-1.1$ and $P_{ 0 }=0.2$. Optical polarimetry data from \citep{aguifernandez_varying_2024}.}
    \label{fig:GRB210610B}
\end{figure}

A long-duration burst at $z=1.134$ was detected by Fermi ($T_{90}= 69$\,sec) and Swift ($T_{90}=55$\,sec). The optical and X-ray light curves are modelled using a broken power-law. The optical light curve remains initially flat until the break at $\sim 7.8$ hours, after which it steepens to a decay index of 1.85. The X-ray light curve appears to undergo a simultaneous break and shares the same post break decay index, though its pre-break decay is faster than that of the optical counterpart \citep{aguifernandez_varying_2024}. 
Polarimetric observations were conducted with the Calar Alto Telescope and the VLT. The first measurements, taken at around 2.9 hours,  revealed a relatively  high polarisation degree of $\sim 4.5\%$ as shown in Figure \ref{fig:GRB210610B}. During the second observation period (5.8--6.7 hours), which is close to the break time, the polarisation level dropped significantly to 0.18--0.6\%.  However, after the break, the polarisation degree increased again to $\sim 2\%$ (30.4--30.6 hours). Interestingly, the PPA rotates by 54 degrees between the pre- and post-break measurements (the bottom panel), suggesting a significant shift that deviates notably from the 90 degree rotation expected for axisymmetric jets. 

This event is a strong candidate for modelling with a non-axisymmetric jet. Here, we assume top-hat, non-axisymmetric jets for our analysis. While a structured, non-axisymmetric jet could also explain the event if the line of sight intersects the jet core, the constraint on the jet head's eccentricity remain similar. 
Given the large deviation of $|\Delta \phi-\pi/2|=36$ degrees, and the distinct trough structure $R_{pd} \geq 10$, we performed 5000 random samplings. For a lower eccentricity ($e=0.3$), most cases (87\%) yield $R_{pd} > 10$. However, virtually none satisfy $|\Delta \phi-\pi/2| >$ 36 degrees.  For $e=0.75$, the fraction of cases meeting the first, second and both conditions are 64\%, 35\% and 21\%, respectively.  The lines of sight that satisfy both conditions are characterised by intermediate azimuthal angles: $ 0.4 < \chi < 1.5$ and a relatively central region: $\xi < 0.7$. These findings suggest this event could be explained within a non-axisymmetric jet framework. 
However, there are caveats in the jet-break modelling. 1) The flat optical light curve before the break suggests significant energy injection into the forward shock at early times. Since ejecta from the central engine can carry large-scale magnetic fields \citep{mundell_highly_2013} unlike forward shocks, which generates highly tangled magnetic fields, the polarisation signals before the break may be significantly influenced by emission from refreshed shocks (energy injection).  2) The post-break decay index of the optical afterglow ($1.85\pm0.04$) is close to, but still shallower than, the expected value ($\sim 2$) for the standard fireball model \citep{aguifernandez_varying_2024}.

\section{Conclusions}\label{sec:conc}

We have investigated the afterglow polarisation signals around a jet break for non-axisymmetric jets. For jets with top-hat luminosity density distributions, the polarisation degree curves exhibit two peaks in general, even when the jet head is elliptical. However, a complete polarisation drop between the peaks is generally absent. Unlike axisymmetric jets, where the polarisation position angle (PPA) undergoes a sudden jump at the trough, the PPA in non-axisymmetric jets evolves smoothly. Additionally, the change in PPA between the two peaks can deviate significantly from 90 degrees. In our random line of sight sampling, 50\% of the cases show deviations of more than 18 degrees, and 10\% exceed 31 degrees for $e=0.6$. 

If the jet's luminosity density gradually declines following a power-law profile, the polarisation curves exhibit a single peak. We find the shapes of the curves are similar for $\alpha =2$ and 3.  In this structured jet case, the PPA can show a slight evolution for non-axisymmetric jets, in contrast to a  constant PPA expected in axisymmetric jet models. Over the full width at half maximum of the polarisation peak, the PPA rotates by 16 degrees for $e=0.6$ with a line of sight at $\chi=\pi/4$ and $\xi=2 (\theta_c/\theta_j)$. The results remain consistent for $\alpha=2$ and 3. 

For structured, non-axisymmetric jets, in which a uniform central core is surrounded by an outer region with a power-law luminosity profile, the polarisation signals depends on the line of sight location. This model can account for a wide range of polarisation behaviours,  as discussed in the case study section. When the line of sight intersects the uniform jet core, the signals resemble those of top-hat jets.  If the line of sight passes through the jet head - whether in the core or outer region - near its semi-major or semi-minor axis, the geometry becomes more symmetrical, making the polarisation signals of non-axisymmetric jets similar to those of axisymmetric jets. Significant deviations in the PPA rotation from 90 degrees (or from a constant PPA in the case of a structured jet) arise when the line of sight has an intermediate azimuthal angle ($\chi \sim \pi/4$) and a relatively small offset angle ($\xi \lsim 0.4$), i.e., when it is close to the jet axis.

As a simple toy model for a structured jet, we considered a case where the luminosity density in the jet's outer region follows a power-law profile. While this simplified model should captures the key features of polarisation signals in structured jets, the angular dependence of luminosity density in a more realistic scenario arises from the lateral energy distribution - where the jet core has a higher energy per solid angle, and the outer regions gradually contain less energy per solid angle. This also induces the angular variations in the Lorentz factor, with the core having a higher Lorentz factor and progressively decreasing toward the outer regions. As a result, the visible region is no longer circularly symmetric, toward the jet centre, the relativistic beaming is stronger and luminosity contribution is larger (e.g., Figure 2 in \citealt{beniamini_robust_2022}). The impact of this deformed visible region will be explored in a future paper. 

When constructing the polarisation degree light curves shown in Figures \ref{fig:pcurve}, \ref{fig:structured_outside_core} -\ref{fig:GRB210610B}, we assumed the forward shock dynamics $\Gamma \propto t^{-3/8}$ for the ISM. In the wind medium, the curve shapes are slightly modified due to different forward shock dynamics $\Gamma \propto t^{-1/4}$. However, the overall morphology (e.g. the presence of two peaks or a single peak) remains unchanged and the discussion on $\Delta \phi$ should be similar. 

If the jet head has a highly eccentric shape, it can be characterised by two angular scales: $\theta_{j}$ (the opening angle along the semi-major axis) and $\theta_{j2}$ (along the semi-minor axis).   As long as $\Gamma > 1/\theta_{j2}$, the flux follows the typical blast wave emission evolution, decaying approximately as $t^{-1}$. When $\Gamma \propto t^{-3/8}$ drops below this first critical threshold, photons begin to be emitted over a larger solid angle, increasing from $\theta_{j}\theta_{j2}$ to $\Gamma^{-1}\theta_{j}$, causing the flux to decay as $t^{-11/8}$. As  $\Gamma$ continues to decrease and falls below the 2nd threshold $1/\theta_{j}$, the emission solid angle expands further to $\Gamma^{-2}$, leading to a steeper decay of $t^{-7/4}$. 
Since these breaks arise from geometrical effects, they are expected to be monochromatic, similar to the usual jet break. 
The two jet break times $t_{j}$ and $t_{j2}$, with $t_{j2} < t_j$, satisfy the relation
\begin{eqnarray}
t_{j2}/t_j &\sim& \paren{1-e^2}^{4/3}.
\end{eqnarray}
For axisymmetric jets ($e=0$), it simplifies to the standard single jet break. 
For a mildly deformed jet head with $e=0.6$, the ratio is approximately $t_{j2}/t_j =0.55$. As a result, the breaks will appear as a smooth transition rather than two distinct ones. While we have assumed a top hat-jet in the ISM to estimate the break time ratio, a wind-like medium would further smooth out each break \citep{panaitescu_analytic_2000}, making them appear as single gradual break. Detecting these breaks in a light curve (requiring $t_{j2}/t_j \ll 1$) would allow for an estimate of the jet's eccentricity as 
\begin{eqnarray}
e &\sim& \sqrt{ 1-(t_{j2}/t_j)^{3/4}}. 
\end{eqnarray}
A more detailed discussion of jet breaks in light curves, including estimates for non-axisymmetric structured jets, will be investigated in a future study.  

Although our case study section focused on optical afterglow observations, the jet break is a geometrical effect, meaning our results are applicable across different wavelengths (e.g., X-ray, radio). The number of events with measured radio polarisation has recently increased (\citealt{urata_first_2019}; \citealt{corsi_upper_2018}; \citealt{laskar_alma_2019}; \citealt{urata_simultaneous_2023}). Significant advancement in X-ray polarimetry in recent years (e.g., Astrosat, IXPE, POLAR, XPoSat) suggest that future X-ray satellites  may be capable of detecting and analysing polarisation signals days after GRBs. Since both X-ray and optical bands are expected to lie above the typical frequency of the forward shock emission around jet breaks, the relativistic limb-brightening  effect should be pronounced in both bands \citep{granot_images_1999}. 

We adopted a conventional scenario in which magnetic fields are amplified by plasma kinetic instabilities at the shock, resulting in a coherence length scale comparable to the plasma skin depth scale \citep{medvedev_generation_1999}. Recently, large-scale turbulent magnetic fields driven by magnetohydrodynamic instabilities have also been proposed (e.g., \citealt{kuwata_synchrotron_2023}). In this scenario,  the coherence length scale is comparable to the thickness of the blast wave. In the isotropic field case, both the polarisation degree and the PPA change randomly and continuously over time, resembling the behaviour described in the classical patch shell model (e.g., \citealt{gruzinov_gammaray_1999,nakar_polarization_2004}). By incorporating magnetic field anisotropy and the observer's viewing angle, this model can also account for a variety of polarisation behaviour in afterglows. Notably, in this framework, the degree of radio polarisation can exceed that in the optical band \citep{kuwata_largescale_2024a}.  Simultaneous polarimetric observations across multiple wavelengths  will be crucial for constraining the magnetic field amplification mechanism and improving our understanding of polarisation signals in afterglows. 

\section*{Acknowledgements}
We thank the anonymous referee for their constructive comments
and suggestions,  which have helped to improve the manuscript.
We are also grateful to Gavin P. Lamb for useful discussions.
This research was supported by the Science and Technology Facilities Council (STFC) grants.

\section*
{Data availability}
The data underlying this article will be shared on reasonable request to the corresponding author.


\bibliographystyle{mnras}
\bibliography{main} 



\appendix

\section{Peak finding algorithm}\label{app:pfa}
To identify peaks and their locations in numerical polarisation curves, we use a peak-finding algorithm similar to the one discussion in  \citealt{li_lognormal_1996}. 
Originally developed for identifying peaks in prompt gamma-ray light curves, this method is well-suited for our analysis.  The detailed procedure is as follows: 
(1) A numerical polarisation curve consists of discrete data points, representing the polarisation degree [\%] as a function of time. A data point is considered a candidate peak if its value is higher than that of its neighbouring points on both sides. The candidate peak has a polarisation degree $P_{p}$ [\%] at time $t_p$. 
(2) For each candidate peak, we search both sides to identify data points with polarisation degree $P_1$ at $t=t_1<t_p$ and $P_2$ at $t=t_2>t_p$ that satisfy the conditions:
\begin{eqnarray}\label{eq:pfa}
    P_p - P_{i} &\ge& N_v \sqrt{P_p} 
\end{eqnarray}
where $i=$ 1,2 and $N_v=5$. 
(3) The search stops when either: both $P_1$ and $P_2$ are found, confirming $P_p$ as a true peak, or a polarisation degree higher than $P_p$ is encountered before finding a data point satisfying Eq. \eqref{eq:pfa} on either side of $t_p$, in which case $P_p$ is discarded as a false peak. 

After this step, all the peaks (one or two peaks in our case) should have been identified. If two peaks are found, the minimum point between them is designated as a trough. This method effectively identifies peaks in polarisation curves while mitigating the effects of small fluctuations due to numerical errors, including the segmentation of the jet head.


\bsp	
\label{lastpage}
\end{document}